\documentclass{article}
\usepackage{spconf,amsmath,graphicx}
\usepackage{diagbox}
\usepackage{threeparttable}
\usepackage{booktabs}
\usepackage{multirow,comment}
\usepackage{bm}
\usepackage{amsfonts,color}
\usepackage{mathtools}

\title{CycleFlow: Purify Information Factors by Cycle Loss}
\name{Haoran Sun, Chen Chen, Lantian Li$^*$, Dong Wang$^*$}
\address{Center for Speech and Language Technologies, BNRist, Tsinghua University, China\\
Department of Computer Science and Technology, Tsinghua University, China}
\begin{document}
%
\maketitle
\begin{abstract}

SpeechFlow is a powerful factorization model based on information
bottleneck (IB), and its effectiveness has been reported by several studies.
A potential problem of SpeechFlow, however, is that if the IB channels are not well
designed, the resultant factors cannot be well disentangled. In this study,
we propose a CycleFlow model that combines random factor substitution and cycle loss
to solve this problem. Experiments on voice conversion tasks demonstrate that
this simple technique can effectively reduce mutual information among individual factors, 
and produce clearly better conversion than the IB-based SpeechFlow.
CycleFlow can also be used as a powerful tool for speech editing.
We demonstrate this usage by an emotion perception experiment.

\end{abstract}

\section{Introduction}
\label{sec:intro}

A distinct talent of human beings is that we can decompose the information
load within speech signals accurately and effectively~\cite{picton1971human,werner2019human}.
Researchers have made quite some attempts to simulate this
ability and presented a couple of information factorization models to decompose
speech~\cite{pell2011emotional,shechtman2019sequence,chou2018multi,chou2019one}.

Most of the current speech factorization models are based on deep neural nets.
Early research employed supervision
from several tasks and used it to factorize speech information in a cascade way~\cite{li2018deep}.
Later research mostly focused on unsupervised learning. For example,
Hsu et al.~\cite{hsu2017unsupervised,hsu2018hierarchical} tried to
establish statistical models in the latent space and
factorize speech signals according to the different statistical properties of
the target factors.

Recently, Qian et al.~\cite{qian2019autovc,qian2020unsupervised} presented
a factorization model based on information bottleneck (IB)~\cite{tishby2000information}, and
reported good performance in voice conversion (VC) tasks. The basic idea behind the
IB-based approach is that by designing appropriate IB channels, one can
control the type and amount of the information flow.
AutoVC is the first successful IB-based VC model~\cite{qian2019autovc}. The architecture is
a conditional autoencoder (AE) where speaker identity is used as condition.
As the speaker information has been known by the model,
the latent code only represents the content, leading to speaker-content decomposition.
SpeechFlow is an extension of AutoVC and represents more factorization capability~\cite{qian2020unsupervised}.
Using carefully designed IB channels,  the model can factorize
speech signals into fine-grained factors, including timbre, rhythm, pitch and content.

In spite of the promising results,
the design of the IB channels is difficult within SpeechFlow, and an improper setting (e.g., dimension of factors
or the resampling scheme) inevitably leads unsatisfied factorization. Unfortunately, discovering an optimal setting
for the model is often a difficult task.
This problem has been noticed by the original authors
and some other researchers~\cite{wang2021adversarially,qian2021global}.

In this study, we present a novel model CycleFlow to purify the information factors.
The main design is a random factor substitution (RFS) operation and a cycle loss.
We highlight that this technique is simple and general and can be applied to any factorization model, although 
in this study we test it with the SpeechFlow architecture, hence the name CycleFlow.
We show theoretically and empirically that the new design indeed
produces more disentangled factors. Moreover, CycleFlow can be used as a powerful tool for
speech editing, hence useful for broad research fields.
We demonstrate the usage by an emotion perception experiment.


\section{Method}
\label{sec:method}

In this section, we revisit the SpeechFlow model presented in~\cite{qian2020unsupervised},
then propose our contribution.
The entire architecture is illustrated in Fig.~\ref{fig:impro}.

\subsection{Revisit SpeechFlow}

SpeechFlow decomposes speech signals into four separate informational factors:
rhythm $\bm{Z}_r$, pitch $\bm{Z}_f$, content $\bm{Z}_c$ and timbre $\bm{Z}_t$ by four encoders:
\begin{equation}
 \begin{aligned}
  &\bm{Z}_r  = \bm{E}_r(\bm{S}),              \nonumber \\
  &\bm{Z}_f  = \bm{E}_f(RR(\bm{P})),   \nonumber \\
  &\bm{Z}_c =  \bm{E}_c(RR(\bm{S})),   \nonumber \\
  &\bm{Z}_t =  \bm{E}_t(\bm{S}),              \nonumber
 \end{aligned}
\end{equation}
\noindent where $RR$ denotes \emph{random resampling} that will be explained below.
Note that the input of the pitch encoder is pitch contour $\bm{P}$, rather than speech $\bm{S}$.

The training objective of the model is to reconstruct the original speech $\bm{S}$ from these information factors:
\begin{equation}
\label{eq:rec-loss}
 \mathcal{L}_{rec} = ||~\hat{\bm{S}} - \bm{S}~||^2,
\end{equation}
\noindent where $||\cdot||$ denotes the $\ell_2$-norm. $\hat{\bm{S}}$ is the reconstructed speech, which is obtained by
a decoding process:
\begin{equation}
 \hat{\bm{S}} = \bm{D}(\bm{Z}_r, \bm{Z}_f, \bm{Z}_c, \bm{Z}_t)~.
\end{equation}

To ensure that the information factors hold their desired information after model training,
the encoders need some special designs, motivated by the IB theory.
The first design choice is to select appropriate dimensions for the four factors so as to ensure minimum sufficient capacity.
Secondly, the timbre factor is specified (not need to learn) by a deep speaker vector~\cite{variani2014deep}.
Thirdly, a random resampling operation ($RR$) shrinks or stretches the duration of the inputs to
the pitch and content encoders in order to prevent them from learning rhythm information.
With these designs, different information tend to be captured by different
factors, as shown in~\cite{qian2020unsupervised}.

\subsection{Our proposed CycleFlow}
\label{sub:impro}

Although the design of SpeechFlow is exquisite, in practice, there is no guarantee
that the encoders extract disentangled factors.
This is because the
IB channels, i.e., the dimensions of the information factors are difficult to choose,
and the resampling process is also not easy to control.
As we will show in Section~\ref{sec:rel}, high mutual information (MI) often exists between the
factors produced by SpeechFlow.

\begin{figure}[h]
	\centering
	\includegraphics[width=1\linewidth]{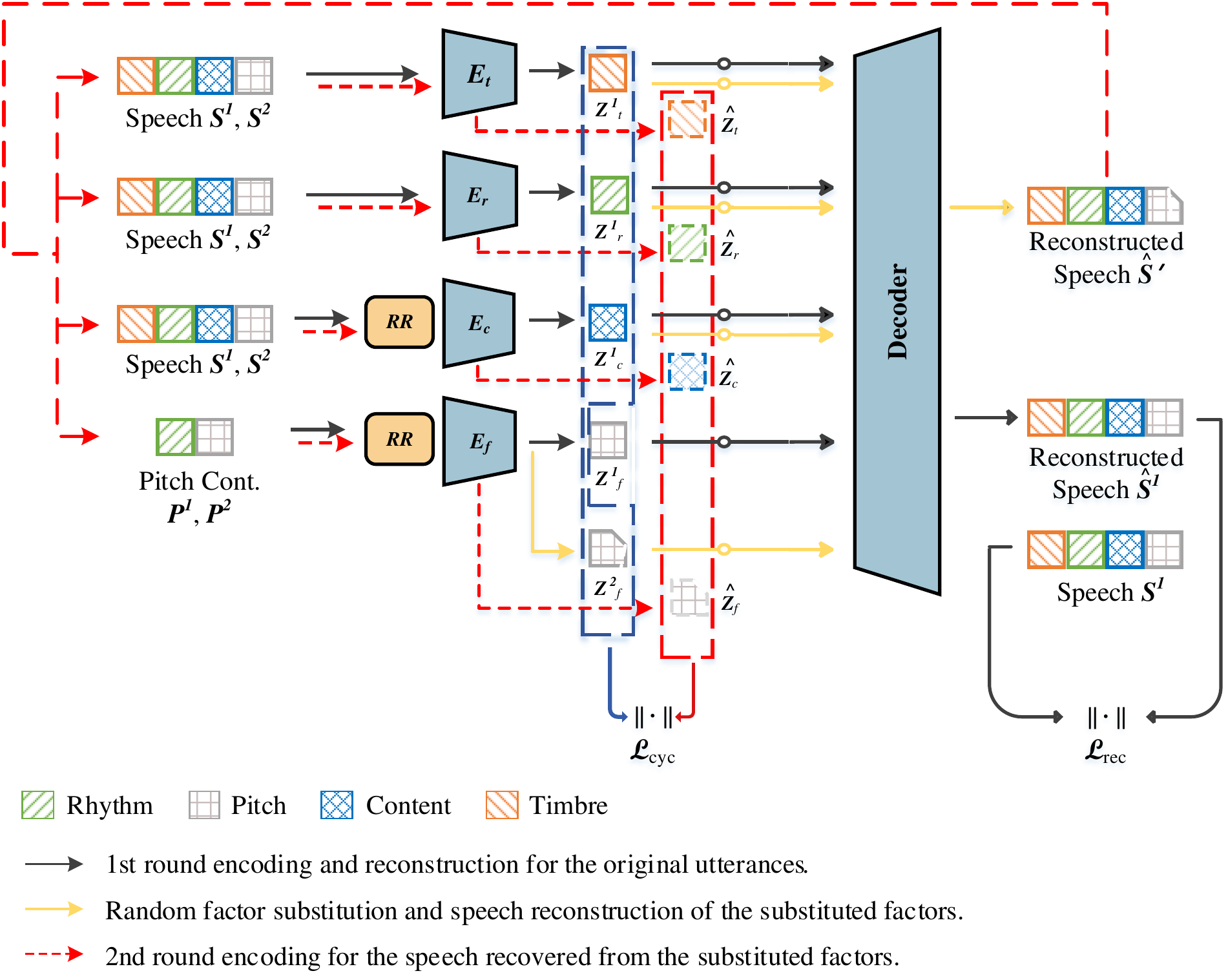}
	\caption{The architecture of CycleFlow.}
	\label{fig:impro}
\end{figure}


CycleFlow is proposed to purify the information factors by introducing a cycle loss.
The idea of cycle loss is motivated by CycleGAN~\cite{zhu2017unpaired},
but here we combine the cycle loss with a factor substitution in order to
encourage information disentanglement.
The new design is shown as yellow solid lines, red dash lines and blocks in Fig.~\ref{fig:impro}.

To make the presentation clear, we use two utterances, denoted by $\bm{S}^1$ and $\bm{S}^2$, to demonstrate
the computing process as follows.

\begin{itemize}
\item {\bf 1st round encoding}: Firstly encode $S^1$ and $S^2$, resulting in two sets of factors:
$\bm{Z}^1 = \{\bm{Z}^1_r, \bm{Z}^1_f, \bm{Z}^1_c, \bm{Z}^1_t\}$
and
$\bm{Z}^2=\{\bm{Z}^2_r, \bm{Z}^2_f, \bm{Z}^2_c, \bm{Z}^2_t\}$.

\item {\bf Random factor substitution (RFS)}: Randomly choose a factor from $\bm{Z}^2$,
and use it to replace the corresponding factor in $\bm{Z}^1$.
Suppose that the selected factor is $\bm{Z}^2_f$, we get a new factor set
$\bm{Z}' = \{\bm{Z}^1_r, \bm{Z}^2_f, \bm{Z}^1_c, \bm{Z}^1_t\}$.

\item {\bf Speech reconstruction}: Forward $\bm{Z}'$ to the decoder and produce the reconstructed speech $\hat{\bm{S}}'$.

\item {\bf 2nd round encoding}: Encode $\hat{\bm{S}}'$ and obtain
 $\hat{\bm{Z}}' = \{ \hat{\bm{Z}}_r', \hat{\bm{Z}}_f', \hat{\bm{Z}}_c', \hat{\bm{Z}}_t' \}$.

\item  {\bf Cycle loss computation}:  The cycle loss is computed as follows:

\begin{equation}
\label{eq:cyc}
\mathcal{L}_{cyc} =  ||\bm{Z}'-\hat{\bm{Z}}'||^2.
\end{equation}

\end{itemize}

The final loss combines the reconstruction loss shown in Eq.(\ref{eq:rec-loss}) and the cycle loss shown in Eq.(\ref{eq:cyc}):

\begin{equation}
\mathcal{L}   = \mathcal{L}_{rec}   +   \alpha *	\mathcal{L}_{cyc}~,
\end{equation}
\noindent where $\alpha$ is a hyperparaemter and is empirically set to be 5.0 in our experiments.


\subsection{Theoretical analysis on CycleFlow}
\label{sub:expla}

We show that under some moderate conditions, the cycle loss combined with RFS
theoretically improves information disentanglement.

Let's define $\bm{Z}'=\{\bm{Z}'_1,\bm{Z}'_2\}$  the factors after RFS.
Now reconstruct $\hat{\bm{X}}'$ from $\bm{Z}'$ and conduct the 2nd round encoding to get
$\hat{\bm{Z}}'= \{ \hat{\bm{Z}}'_1, \hat{\bm{Z}}'_2\}$. Our purpose is to let $\hat{\bm{Z}}_1'$ \emph{fully} determined by $\bm{Z}'_1$
and $\hat{\bm{Z}}_2'$ \emph{fully} determined by $\bm{Z}'_2$. This can be obtained by minimizing
the conditional entropy $H(\hat{\bm{Z}}_1'|\bm{Z}'_1)$ and $H(\hat{\bm{Z}}_2'|\bm{Z}'_2)$, formulated by the following objective:

\begin{eqnarray}
\mathcal{L}_{ch}&=&H(\hat{\bm{Z}}_1'|\bm{Z}'_1)+ H(\hat{\bm{Z}}_2'|\bm{Z}'_2) \nonumber \\
&=&-\mathbb{E}_{\bm{Z}_1'} \mathbb{E}_{\bm{Z}_2'}\log p(\hat{\bm{Z}_1'}|\bm{Z}'_1) - \mathbb{E}_{\bm{Z}_2'} \mathbb{E}_{\bm{Z}_1'} \log p(\hat{\bm{Z}_2'}|\bm{Z}'_2)~,  \nonumber
\end{eqnarray}

\noindent where the inner expectation specifies the source of the conditional variation. If we further assume that the
conditional probabilities $p(\hat{\bm{Z}_1'}|\bm{Z}'_1)$ and $p(\hat{\bm{Z}_2'}|\bm{Z}'_2)$
are isotropic Gaussian with variation $\sigma$, then we have:

\begin{equation}
\begin{aligned}
\mathcal{L}_{ch}&= \mathbb{E}_{\bm{Z}_1'} \mathbb{E}_{\bm{Z}_2'} ||\hat{\bm{Z}_1'} - \bm{Z}'_1||_2^2 + \mathbb{E}_{\bm{Z}_2'} \mathbb{E}_{\bm{Z}_1'} ||\hat{\bm{Z}_2'} - \bm{Z}'_2||_2^2 + C(\sigma) \nonumber \\
&\propto\mathbb{E}_{\bm{Z}'}   ||\hat{\bm{Z}'} - \bm{Z}'||_2^2~, \nonumber
\end{aligned}
\end{equation}
\noindent where $C(\sigma)$ is a constant depending on $\sigma$. This is just the cycle loss shown in Eq.(\ref{eq:cyc}).

Further notice that $\bm{Z}_1'$ and $\bm{Z}_2'$ are independent,
and in the case $\mathcal{L}_{ch}=0$, $\hat{\bm{Z}}_1'$ is totally dependent on ${\bm{Z}}_1'$, and $\hat{\bm{Z}}_2'$ is totally dependent on ${\bm{Z}}_2'$,
we have:

\begin{eqnarray}
p(\hat{\bm{Z}}'_1|\hat{\bm{Z}}'_2) &=& \int_{\bm{Z}'_1, \bm{Z}'_2} p(\hat{\bm{Z}}'_1|\bm{Z}'_1, \bm{Z}'_2, \hat{\bm{Z}}'_2) p(\bm{Z}'_1,\bm{Z}'_2|\hat{\bm{Z}}'_2)  \nonumber \\
&=& \int_{\bm{Z}'_1,\bm{Z}'_2} p(\hat{\bm{Z}}'_1|\bm{Z}'_1) \frac{p(\hat{\bm{Z}}_2'|\bm{Z}'_1,\bm{Z}'_2) p(\bm{Z}'_1)p(\bm{Z}'_2)}{p(\hat{\bm{Z}}_2')} \nonumber \\
&=& \int_{\bm{Z}'_1} p(\hat{\bm{Z}}'_1|\bm{Z}'_1) \int_{\bm{Z}'_2}\frac{p(\hat{\bm{Z}}_2'|\bm{Z}'_2) p(\bm{Z}'_2)}{p(\hat{\bm{Z}}_2')} \nonumber \\
&=&   p(\hat{\bm{Z}}'_1)~. \nonumber
\end{eqnarray}

\noindent This result means that if the model has been well trained and the cycle loss converges to zero, then the factors are mutually independent. Note that our derivation
does not rely any IB design, hence equally applicable to IB-free models.

We highlight that the above result does not imply that real speech signals will be factorized in a disentangled way, as the reconstructed speech from the RFS factors,
which is indeed decomposed into disentangled factors by the 2nd round encoding, does not necessarily represent real speech. However, we can enforce
 the representation by (1) Use factors produced from real speech to perform RFS; (2) Combine the cycle loss and the reconstruction loss.
The training process presented in the previous section reflects these theoretical concerns.

\section{Related works}
\label{sec:rel}

Cycle loss in this paper was motivated by the cycle consistency loss in CycleGAN~\cite{zhu2017unpaired}.
The cycle loss has been proved to be useful in domain adaptation~\cite{hoffman2018cycada},
image-to-image translation~\cite{zhu2017multimodal}, medical image synthesis~\cite{zhang2018translating}
and voice conversion~\cite{lee2020many}.
We borrowed the idea from these studies and combined it with RFS
to improve information disentanglement in speech factorization models.
We also noticed that some authors employed adversarial training to pursue the same goal~\cite{wang2021adversarially},
but adversarial training is notoriously unstable and the computation cost is significant.

\section{Experiments}
\label{sec:rel}

We tested the proposed CycleFlow on some quantitative and qualitative experiments.
In voice conversion tasks, we tried to convert speaking style, speaking timber, and both.
The source code and audio samples are available at \emph{http://cycleflow.cslt.org}.

\subsection{Data and configurations}
\label{sub:data}
The corpus used to implement our experiments is the CSTR VCTK corpus~\cite{veaux2017cstr}.
Speech signals from 100 speakers were used to conduct model training,
and those from 8 speakers were used to perform testing. The speakers and utterances in the training and test
sets are not overlapped. All the speech signals were uniformly formatted to
16kHz, 16-bits, and then were transformed to spectrograms
no longer than 192 frames by a short time Fourier transform (STFT).

SpeechFlow was implemented
using the source code published online\footnote{https://github.com/FantSun/Speechflow.}, and the same 
code was adapted to implement CycleFlow. We mostly followed
the settings in the original repository, including network structure, data processing steps, and training scheme.
The only major modification was that we changed the input to the timber encoder from one-hot speaker codes to continuous
speaker vectors~\cite{variani2014deep} in order to improve generalization capability.

\subsection{Mutual information}
\label{sub:mi}

In the first experiment, we investigate the mutual information (1) between the original speech and the encoded factors,
and (2) between pairs of factors. Ideally, if the reconstruction loss is comparable (which is the case in our experiments), we hope all
these mutual information is as low as possible.

A difficulty is that all the concerned variables are continuous and their dimensions are different.
To tackle the problem,
we firstly clustered each variable into 10 classes by K-means clustering,
and then calculated mutual information based on the cluster IDs of the variables.

The results are shown in Table \ref{tab:mi}. It can be observed that in all the comparisons,
CycleFlow achieves less mutual information.
For rows 1, 2 and 3, this means that all the encoders eliminate more irrelevant information from the original speech.
For row 4, 5 and 6, this represents that the factors are more independent. All these results demonstrate that
CycleFlow can produce more disentangled factors.



\begin{table}[!htp]
 \caption{Mutual information between signals and factors, and factor pairs. $\bm{S}$ denotes the original
 speech signal. $\bm{Z}_c$, $\bm{Z}_r$, $\bm{Z}_f$ denote content, rhythm and pitch factors.}
 \centering
 \begin{tabular}{lccc}
   \toprule
   \multirow{1}{*}{No.} & \multicolumn{1}{c}{Factors} & \multicolumn{1}{c}{SpeechFlow} & \multicolumn{1}{c}{CycleFlow} \\
   \cmidrule(r){1-1}     \cmidrule(r){2-2}  \cmidrule(r){3-3}  \cmidrule(r){4-4}
   1  &  $\bm{S}$ vs. $\bm{Z}_c$   & 0.5093   & \textbf{0.3659}  \\
   2  &  $\bm{S}$ vs. $\bm{Z}_r$   & 0.7133   & \textbf{0.6155}  \\
   3  &  $\bm{S}$ vs. $\bm{Z}_f$   & 0.5515   & \textbf{0.4312}  \\
   4  &  $\bm{Z}_c$ vs. $\bm{Z}_r$   & 0.5155   & \textbf{0.3400}  \\
   5  &  $\bm{Z}_c$ vs. $\bm{Z}_f$   & 0.4455   & \textbf{0.2983}  \\
   6  &  $\bm{Z}_r$ vs. $\bm{Z}_f$   & 0.5287   & \textbf{0.4953}  \\
   \bottomrule
\label{tab:mi}
\end{tabular}
\vspace{-5mm}
\end{table}

\subsection{Voice conversion}
\label{sub:subj}

In the second experiment, we conducted voice conversion using SpeechFlow and CycleFlow.
We hired 15 Chinese listeners, each being assigned 70 utterances in total, divided
into 4 test groups: (1) reconstruction quality; (2) speaking style transfer; (3) timber maintenance in style transfer;
(4) speaker conversion with both style and timber transfer.
For each evaluation, we presented listeners two utterances produced by SpeechFlow and CycleFlow respectively,
and asked them to select which one is better according to the specified quality metric.

\begin{figure}[h]
	\centering
	\includegraphics[width=0.6\linewidth]{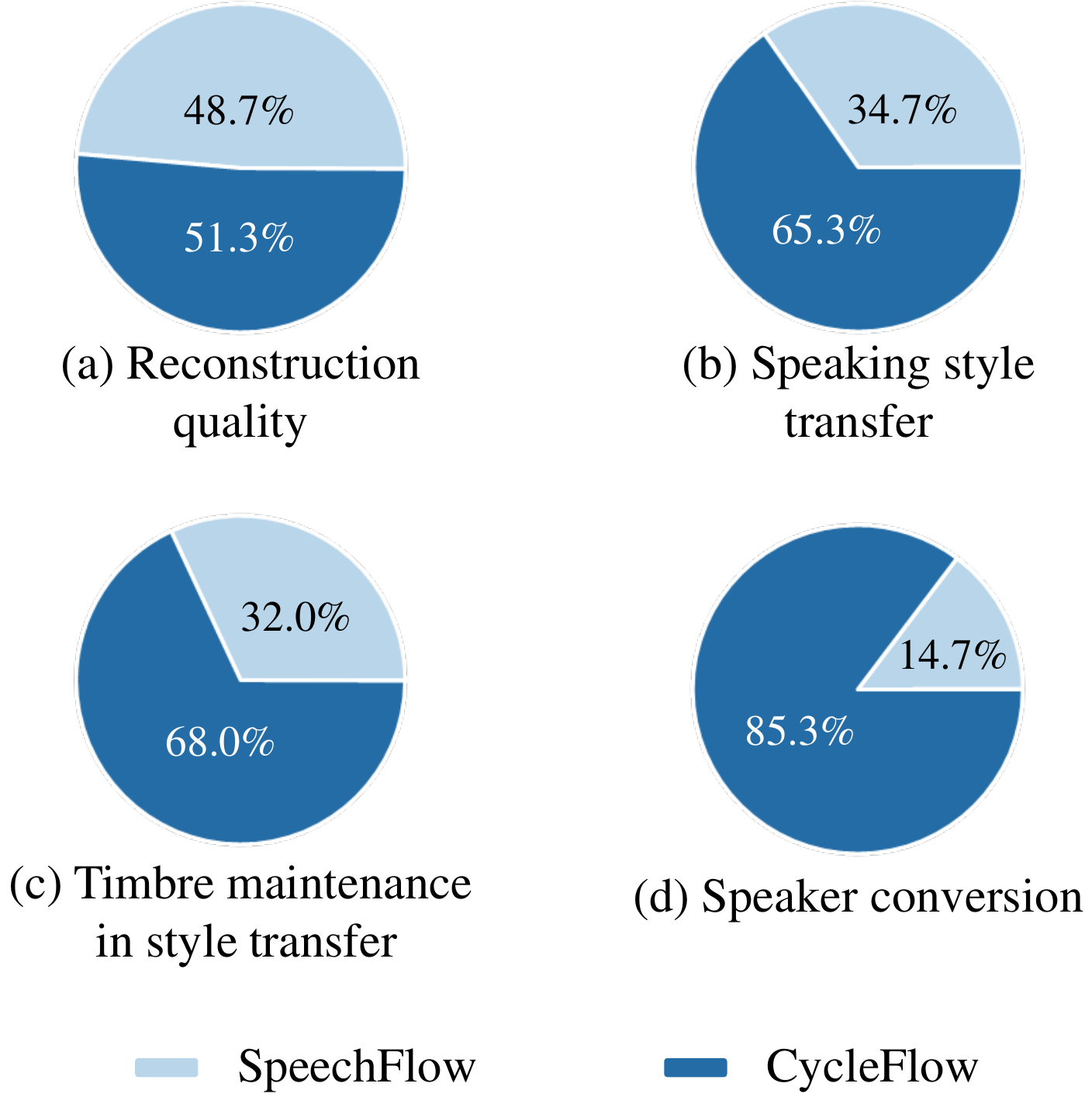}
	\caption{Results of subjective evaluations.}
	\label{fig:pie}
\end{figure}

Fig.~\ref{fig:pie} (a) shows the result of the reconstruction quality test, where
listeners were required to select the reconstructed speech that sounded more similar
to the original one and with better quality.
The result shows that CycleFlow can produce reconstructed speech with comparable quality as SpeechFlow.

Fig.~\ref{fig:pie} (b) shows the result of speaking style transfer, where the
speaking style (including both rhythm and pitch)
of the source speech was changed to that of the target speech. The listeners were
asked to select which utterance gave a better transfer.
The result shows that CycleFlow is significantly better than SpeechFlow.

Fig.~\ref{fig:pie} (c) shows the result if the speaker timber was
maintained when the speaking style was transferred. The experiment
is the same as in Fig.~\ref{fig:pie} (b), except that
the listeners were asked to focus on timber rather than speaking style.
The result shows that CycleFlow clearly wins, and indicates
that it maintains timber better when speaking style is changed.

Fig.~\ref{fig:pie} (d) shows the result of speaker conversion, where both speaking style and timber
was changed. The listeners were asked to select which utterance performed a better conversion
in terms of speaker similarity. Again, CycleFlow shows clear advantage.

As a summary, we demonstrated better independence of speaking style in (b) and (c), and the success of (d)
further demonstrated better independence of speaker timber. Overall, this experiment demonstrated
that more independence was obtained for all the factors by our proposed CycleFlow model:
style (including pitch and rhythm), speaker timber and content. Note that we did not intend to
test pitch and rhythm separately, as it seems not easy for human listeners to identify them individually.

\subsection{Human emotion perception}
\label{sub:emo}

CycleFlow can be used as a speech editing tool. 
In this experiment, we demonstrate the usage using an emotion perception
task, which aims to test if human ears can discriminate different types of
emotion by using only cues of rhythm and pitch.

\begin{figure}[h]
	\centering
	\includegraphics[width=0.7\linewidth]{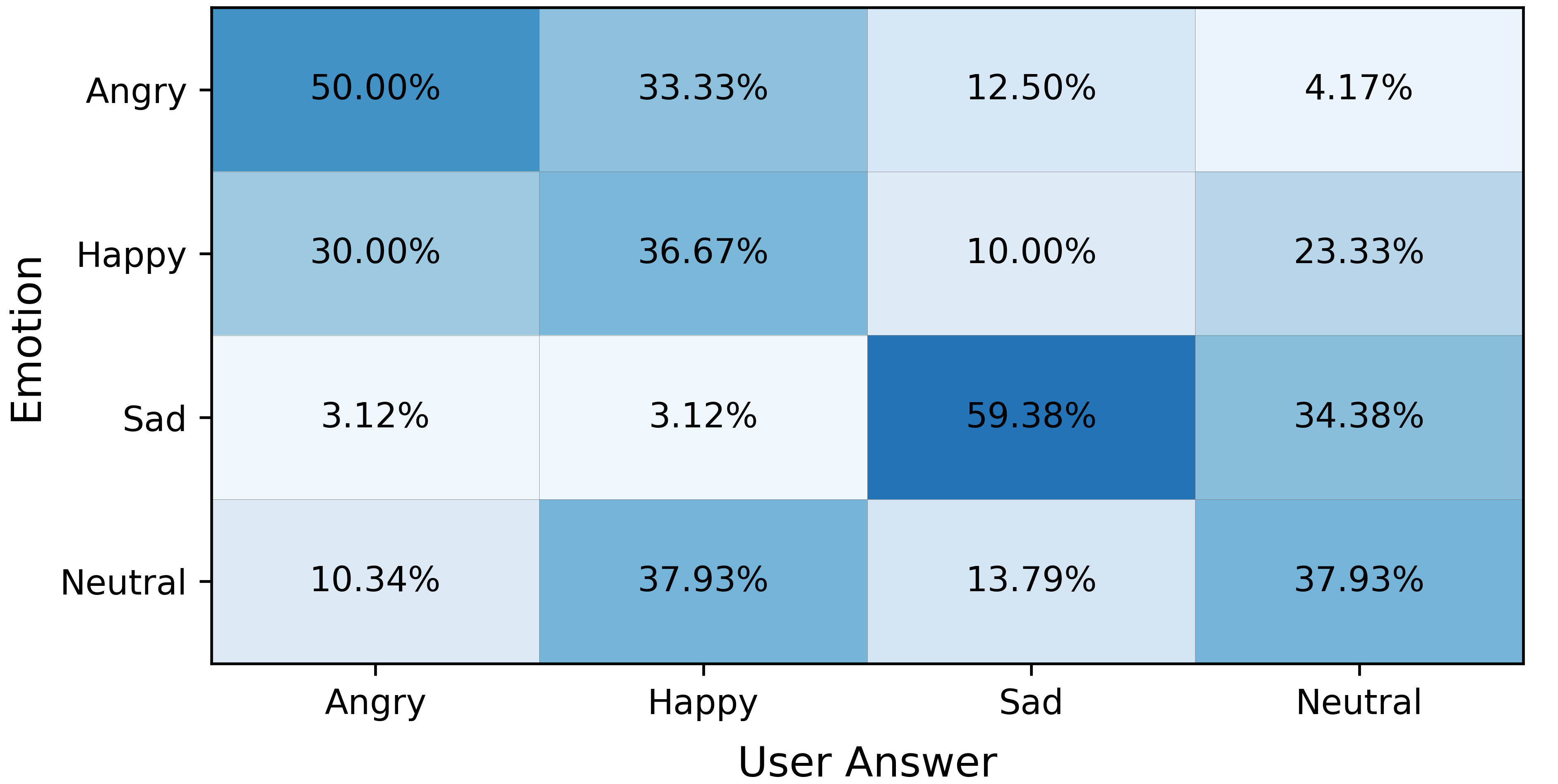}
	\caption{Pair-wised accuracy of human emotion perception test with only rhythm and pitch cues.}
	\label{fig:RF}
\end{figure}

To achieve that goal, we first used CycleFlow to factorize speech signals and
then set the content and timber factor to be constant, and reconstructed the speech.
By this operation, only speaking style (including pitch and rhythm) was retained.
We then hired $8$ listeners to identify the emotion of utterances in four emotion types:
\emph{Angry}, \emph{Happy}, \emph{Neutral}, \emph{Sad}.  $97$ utterances from IEMOCAP~\cite{busso2008iemocap}
were used in the test.
The pair-wised accuracy is shown in Fig.~\ref{fig:RF}.
As can be seen, only with speaking style,
human can tell the emotion of a speech with a good accuracy.
In particular, we seem more excel at identifying angry and sad emotion.
This coincides with our daily experience.

Note that this experiment should not be regarded as a strict scientific study, but a
simple demo of usage.
Moreover, CycleFlow is more suitable than SpeechFlow in this usage, as it can decompose speech 
signals into more disentangled factors.





\section{Conclusion}
\label{sec:conl}

We proposed a novel CycleFlow model for speech information factorization. 
The core of the design is a combination of  factor substitution and cycle loss. 
We demonstrated theoretically and empirically
that the proposed technique can significantly reduce mutual information among factors, and produce
better performance in voice conversion. Moreover, CycleFlow can be used as a
powerful tool for speech editing, hence lending itself to a wide range of scientific research.
Future work involves applying cycle loss to other factorization models, and using it to design 
more detailed factorization. 

\newpage
\bibliographystyle{IEEEbib}
\bibliography{refs}

\end{document}